\documentclass[a4paper]{cas-dc}
\usepackage{amsmath}
\usepackage{graphicx}
\usepackage{tabularray}
\usepackage{eurosym}
\usepackage{url}
\usepackage{lipsum}
\usepackage{soul}
\usepackage{hyperref}
\usepackage[authoryear]{natbib}
\def\tsc#1{\csdef{#1}{\textsc{\lowercase{#1}}\xspace}}
\tsc{WGM}
\tsc{QE}
\tsc{EP}
\tsc{PMS}
\tsc{BEC}
\tsc{DE}

\begin{document}
\let\WriteBookmarks\relax
\def\floatpagepagefraction{1}
\def\textpagefraction{.001}
\newcommand{\fig}[1]{\color{blue}Fig.~\ref{#1}\color{black}}
\shorttitle{Spatiotemporal variability of ride-pooling potential - half a year New York City experiment}

\newcommand{\R}{}
\newcommand{\CC}[1]{\textcolor{black}{#1}}
\newcommand{\CI}[1]{\textcolor{black}{\textit{#1}}}
\newcommand{\CR}[1]{\textcolor{gray}{#1}}

\shortauthors{Olha Shulika et al.}

\title[mode = title]{Spatiotemporal variability of ride-pooling potential – half a year New York City experiment}                      


%


\author[1]{Olha Shulika}[orcid=0000-0002-1912-1115]


\cormark[1]
\ead{olha.shulika@uj.edu.pl}



\affiliation [1] {organization={Faculty of Mathematics and Computer Science, Jagiellonian University},
    addressline={Profesora Stanisława Łojasiewicza 6}, 
    city={Krakow},
    postcode={30-348}, 
    country={Poland}}
\author [2]
{Michal Bujak}
\ead{michal.bujak@doctoral.uj.edu.pl}
\affiliation [2] {organization={Doctoral School of Exact and Natural Sciences, Jagiellonian University},
    addressline={Profesora Stanisława Łojasiewicza 11}, 
    city={Krakow},
    postcode={30-348}, 
    country={Poland}}

\author [2]
{Farnoud Ghasemi}
\ead{farnoud.ghasemi@doctoral.uj.edu.pl}

\author[1]
{Rafal Kucharski}
[orcid=0000-0002-9767-8883]
\ead{rafal.kucharski@uj.edu.pl}
\ead[URL]{https://rafalkucharskipk.github.io/}

\cortext[cor1]{Corresponding author}



\begin{abstract}
Ride-pooling systems, despite being an appealing urban mobility mode, still struggle to gain momentum. 
While we know the significance of critical mass in reaching system sustainability, less is known about the spatiotemporal patterns of system performance. Here, we use 1.5 million NYC taxi trips (sampled over a six-month period) and experiment to understand how well they could  be served with pooled services. We use a utility-driven ride-pooling algorithm and observe the pooling potential with six performance indicators: mileage reductions, travellers' utility gains, share of pooled rides, occupancy, detours, and potential fleet reduction. We report distributions and temporal profiles of about 35 thousand experiments covering weekdays, weekends, evenings, mornings, and nights. We report complex spatial patterns, with gains concentrated in the core of the network and costs concentrated on the peripheries. The greatest potential shifts from the North in the morning to the Central and South in the afternoon. Offering pooled rides at the fare $32\%$ lower than private ride-hailing seems to be sufficient to attract pooling yet dynamically adjusting it to the demand level and spatial pattern may be efficient. The patterns observed in NYC were replicated on smaller datasets in Chicago and Washington, DC, the occupancy grows with the demand with similar trends.
\end{abstract}



\maketitle

\section{Introduction}
Ride-pooling (also called shared ride-hailing or ride-splitting) is a service in which passengers with similar origins or destinations are pooled to the same vehicle to travel together. The total mileage is then reduced and the travel costs can be now shared by the co-travellers \citep{santi2014quantifying,alonso2017demand}. Ride-pooling hardly reached the critical mass in the pre-pandemic period and almost completely stopped during the pandemic \citep{foljanty2022}, nonetheless it remains a promising emerging mobility mode with the significant potential to contribute towards sustainability transitions. Effective pooling can complement public transport \citep{cats2022beyond}, and when demand levels exceed the so-called critical mass, this becomes efficient for all parties: for travellers who pay less, for drivers who can earn more, for the service provider (mobility platform) that can reduce the total mileage and better utilise its fleet, and for the city that can reduce traffic congestion and externalities \citep{shaheen2016shared}.

\begin{figure*}[h!]
	\centering
		\includegraphics[width=\linewidth]{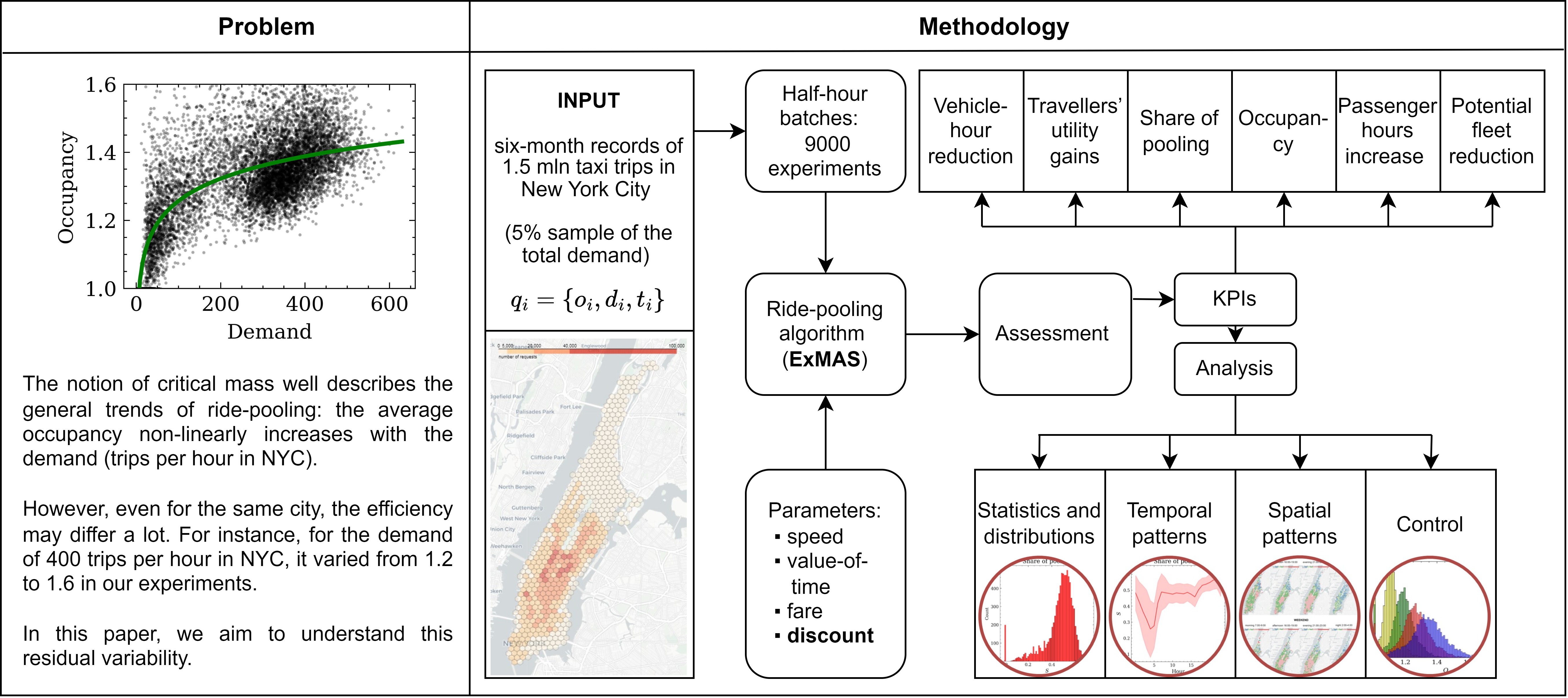}
	\caption{The problem statement (left) and an overview of the methodology (right) applied in this study to understand the potential of ride-pooling.}
	\label{FIG:1}    
\end{figure*}

High demand levels are substantial in reaching the full potential of pooling services, but can be insufficient. The actual performance depends also on operating area, supply concentration, topology etc. \citep{fielbaum2023economies,liu2023scale,manik2020topology}. As a result, identifying the efficiency and attractiveness of ride-pooling for various parties involved is more nuanced and calls for a detailed analysis. While the general trends of shareability have been widely reported, so far, the detailed spatiotemporal analysis was missing. 
 To illustrate the significance of more detailed analysis, let's analyse the following glimpse of our results in \fig{FIG:1}~left. The average occupancy non-linearly increases with the demand level in NYC and the logarithmic elbow curve well explains the general trend, yet the variability among individual experiments (dots) remains huge, e.g. the occupancy for demand levels of 300 trips per hour vary from 1.2 to 1.6.

In this study we aim to explain this residual variability and explore the potential of ride-pooling on the big dataset. To provide more thorough understanding of ride-pooling potential in dense urban areas, we synthesize the results of 9~000 half-hour ride-pooling experiments in which the sample of 1.5 million actual trip requests from New York are pooled into attractive shared rides (\fig{FIG:1}~right). We provide detailed results that go beyond the classical notion of a critical mass. We observe how shareability changes for various days of the week and hours of the day, and report the temporal and spatial distributions of six various ride-pooling performance indicators. The results can help understand the actual potential of ride-pooling and its limitations. The results can help policymakers and transportation network companies develop policies for the development of sustainable, attractive and commercially successful ride-pooling services. 

\subsection{Literature review}

Ride-pooling is the most effective with high demand \citep{santi2014quantifying, alonso2017demand, tachet2017scaling}. While in public transport, the demand density is a key to reach efficiency, it also varies greatly with spatial
and temporal distributions \citep{fielbaum2016optimal}. Here, we want to see if this is also true for ride-pooling services. 

\cite{santi2014quantifying} propose a quantification system to measure the spatial and temporal compatibility of individual trips based on a shareability network of New York City with a high population and taxi traffic density in a relatively geographically small area. \cite{tachet2017scaling} generalise it and observe similar trends also for: San Francisco, Singapore, and Vienna; further extended by \cite{ke2021data} to Chengdu and Haikou.
Notably, the demand for pooled services is elastic and correlates, e.g. with: the availability of alternative public transport \citep{cats2022beyond}, sociodemographic characteristics of population \citep{abouelela2022characterizing,compostella4034722investigation}, urban structure of the city \citep{li2019characterization}, etc. 
\cite{zwick2022ride} analyse the big pre-pandemic demand dataset of the ride-pooling service MOIA in Hamburg and Hanover from May 2019 to February 2020 and propose regression models predicting ride-pooling demand capturing the impact of the density of workplaces, gastronomy and culture. \cite{hou2020factors} analyse the impact of trip, time, and location on the willingness-to-share for Chicago based on TNC data set (for Via, UberPool and LyftShare services). Using multivariate linear regression, the authors reveal the significance of income level and airport trips on travellers’ willingness to share a trip. \cite{du2022spatial} analyse Chicago ride-hailing data from March 2019 and identify how the spatial patterns of socioeconomic and built environment influence ridesplitting performance during morning and evening peak hours on weekdays and weekends. The authors find an increased demand for pooled trips during weekday morning and evening peak hours, which differs from the findings of an earlier study by \cite{li2019characterization}, who study the spatial and temporal characteristics of solo and pooled rides based on DiDi Chuxing data from November 2016. \cite{li2019characterization} analyse the correlation of pooled rides distribution with the built environment and temporal parameters, they find that ride-pooling demand is higher during non-work hours, for longer distances in the direction from the central part of the city. \cite{compostella4034722investigation} found, based on 5,136 trips conducted by 1,991 travellers in California between November 2018 and November 2019, that individuals' mobility profile is the primary determinant in choosing a pooled trip along Transportation Network Companies (TNC) services usage and population density. {\cite{liu2023scale}; \cite{fielbaum2023economies}; \cite{lehe2021increasing} on the NYC data set analyse ride-pooling scale effects, cost-saving opportunities and cost-increasing factors.

Such variability of the demand calls for the detailed analysis of ride-pooling efficiency, we argue and demonstrate that for the same demand levels in the same area the ride-pooling efficiency can vary dramatically (\fig{FIG:1}~left).
We present an overview of previous approaches to ride-pooling potential evaluations in Table \ref{tbl1}. Since large-scale ride-pooling has barely been seen in practise, most of studies rely on simulations. 
In the seminal work of \cite{santi2014quantifying}, the pooling potential is measured with the two indicators: number of trips that can be successfully shared (shareability) and travel time savings. While this shows general patterns, the perspective of travellers and their utility, as well as service operator benefits, is missing. Other measures include occupancy \citep{alonso2017demand, simonetto2019real}. 
\cite{young2020true} measure the shareability potential for big dataset of ride-hailing trips in Toronto with: number of successfully matched trips (matching propensity) and travel detour time. With the results showing that, first, high-demand areas have a greater likelihood of successful ride-sharing and, secondly, longer wait times before the driver arrives can actually increase the chances of matching shared trips. Finally, shared trips experience only a minor variance in travel detour time.
Using the total cost minimization approach that considers the interests of both users and operators, \cite{militao2021optimal} explore the potential for ride-pooling under different scenarios for vehicle automation. 
\cite{du2022spatial} perform temporal-spatial analysis for Chicago and report the share of pooled trips. \cite{kucharski2020exact} follows utility-based demand-centric approach and measure the shareability potential using occupancy (the ratio of total passenger-hours travelled to vehicle -hours travelled with passengers), vehicle-hour reduction, passenger hours increase and travelers’ utility gains for Amsterdam experiments with up to 4~000 travellers per hour. Here, we use this set of indicators as the one providing the complete picture, and apply it for the big dataset. 
\cite{soza2022shareability} follow a similar direction and use the Exact Matching of Attractive Shared Rides (\emph{ExMAS}) algorithm to explore the impact of spatial travel demand patterns on ride pooling performance for different number of attraction centres, density of destinations around each centre, and trip length distribution, using Amsterdam as an example with a fixed level of synthetic demand.  

 \begin{table*}
\caption{Review of previous ride-pooling potential evaluation studies.}\label{tbl1}
\begin{tabular*}{\tblwidth}{m{2cm}m{4cm}m{2cm}m{2cm}m{5cm}}
\toprule
Work & KPIs & Temporal patterns & Spatial patterns & Dataset\\
\midrule

{\cite{santi2014quantifying}} & {Number of shared trips, Travel time savings} & {+} & {-} & {150 million trips: NYC (Manhattan), 6 months of 2011} \\

{\cite{tachet2017scaling}} & {Number of shared trips} & {+} & {-} & {Over 156 million taxi trips: New York City (in 2011), San Francisco (in 2009), Singapore, and Vienna} \\

{\cite{alonso2017demand}} & {Travel delay, Occupancy} & {+} & {-} & {About 3  million trips: NYC (Manhattan), 05.05.2013-11.05.2013} \\

{\cite{simonetto2019real}} & {Travel delay, Occupancy} & {+} & {+} & {About 3  million trips: NYC (Manhattan), 05.05.2013-11.05.2013} \\

{\cite{li2019characterization}} & {Travel delay, Travel detour} & {+} & {-} & {6.1 million ridesourcing orders, Chengdu (China), 01.11.2016-30.11.2016} \\

{\cite{young2020true}} & {Matching propensity (the number of successfully matched trips), Travel detour} & {+} & {+} & {12 million records of ride-hailing trips: Toronto (Canada), 07.09.2016-31.03.2017} \\

 {\cite{militao2021optimal}} & {Vehicle size, Fleet size} & {+} & {-} & {8.7 million trips for one day in 2011: Munich (Germany)} \\

{\cite{kucharski2020exact}} & {Total vehicle hours, Total passenger hours, Travellers’ utility gains, Occupancy} & {+} & {-} & {241~000 trips per working day  in Amsterdam (Netherlands)}\\

{\cite{soza2022shareability}} & {Total vehicle hours, Total passenger cost, Share of pooled trips, Occupancy} & {+} & {+} & {1~000 trip requests are generated during a one-hour period, Amsterdam (Netherlands)}\\ \midrule

{Our research} & {Total vehicle hours, Total passenger hours, Share of pooling, Travellers utility gains, Occupancy, Fleet size} & {+} & {+} & {1.5 million trips, New York City (Manhattan), 6 months of 2016} \\ \bottomrule

 \end{tabular*}
 \end{table*}

\subsection{Methodological approach and contributions}

To reveal the actual potential of ride-pooling, we need a detailed understanding of its performance. Previous studies either relied on synthetic demand data \citep{soza2022shareability}, short time periods \citep{simonetto2019real}, reported only few measures \citep{du2022spatial, tachet2017scaling, militao2021optimal}, adopted the service-operator's perspective \citep{alonso2017demand}, or their results were just an illustration of the algorithm and not analysed in detail \citep{kucharski2020exact}.


Here, we explicitly focus on the analysis of the spatial and temporal patterns of the potential ride-pooling service performance. We use the big dataset of ride-hailing trips and apply the utility driven ride-pooling algorithm to report a set of six indicators measuring the potential of ride-pooling potential: for different parties involved (travellers, drivers, operators and policymakers) during varying time periods (weekends, weekdays, peak hours and nights) and across the different parts of Manhattan. 
In particular, we:
\begin{itemize}
    \item use the big dataset (1.5 million trips) of the actual ride-hailing travel demand spanning over six months,
    \item apply the demand-driven ride-pooling algorithm to make sure that the pooling alternative remains attractive for travellers,
    \item report the comprehensive set of six performance indicators,
    \item run detailed temporal and spatial analysis of the results, 
    \item visualize spatial distributions of ride-pooling performance,
    \item experiment with varying the discount offered,
    \item to understand how general are our findings, we compare NYC results with Chicago and Washington and observe a similar pattern.

\end{itemize}


\section{Methods and data}

The methodology applied to obtain spatial and temporal insights on the potential performance of ride-pooling is illustrated with \fig{FIG:1}~right.
We first collect a big dataset of trip requests with detailed spatial and temporal references. Then, to understand the potential of ride-pooling, we run the utility-based ride-pooling algorithm ExMAS to match travellers to shared rides. We divide the six months of data into about 9~000 batches of half-hour for each of which we run the ride-pooling algorithm.  
ExMAS first identifies all the feasible and attractive pooling combinations and then matches travellers into attractive shared rides. Since pooled rides identified with ExMAS are strictly attractive for travellers, in the matching we may focus on the system (and operators') perspective and minimise the total vehicle mileage.

Notably, we assume that travellers are rational decision makers and opt only for pooled rides more attractive than private ride-hailing (solo-rides). We obtain the detailed results from which we assess the system performance with a variety of indicators. Performance is observed at the batch level (system-wide performance for all trips requested in 30-minute batch), as well as at the individual level (where performance for individual travellers can be used for spatial analysis). The results (input data, ExMAS scripts, and reproducible results) are stored at the public repository (\cite{shulikaGitHub2022}).

\subsection{Dataset}
We analyse ride-pooling potential with over 1.5 million solo taxi trips recorded in Manhattan (NYC) in the first half of 2016. We used sample of NYC Taxi and ride-hailing trip requests from Manhattan \citep{schneider2015analyzing,schneider2019taxi}. Each record ($q_{i}$) contains information on its origin ($o_{i}$), destination ($d_{i}$) and request time ($t_{i}$); the exact route was not reported. Our baseline is a scenario where all travellers use solo ride-hailing. Then, we assess how much of the ride-hailing can be substituted with ride-pooling such that all travellers have at least the same satisfaction with the service (utility formulas) and the overall vehicle millage is reduced. Travellers who cannot be attractively matched into pooled rides are served with solo rides. We used a 5\% sample of the total demand (ca. 60 million trips per year) as a reasonable reference point for the plausible potential market share of ride-pooling. In the pre-pandemic peaks around $15\%$ of the taxi demand was opting for a pooled services, yet only half was actually pooled (in Toronto, \cite{young2020true}), and the post-pandemic the ride-pooling operations are practically ceased. As we report below, the demand levels and density at 5\% market share allowed to observe a variety of ride-pooling regimes (from hypocritical to hypercritical). The temporal profiles of the demand (reported in \fig{FIG:2})~as well as spatial patterns (\fig{FIG:3})~follow the observed general trends reported, e.g., in \cite{schneider2019taxi}. 

\begin{figure*}
	\centering
		\includegraphics[width=\linewidth]{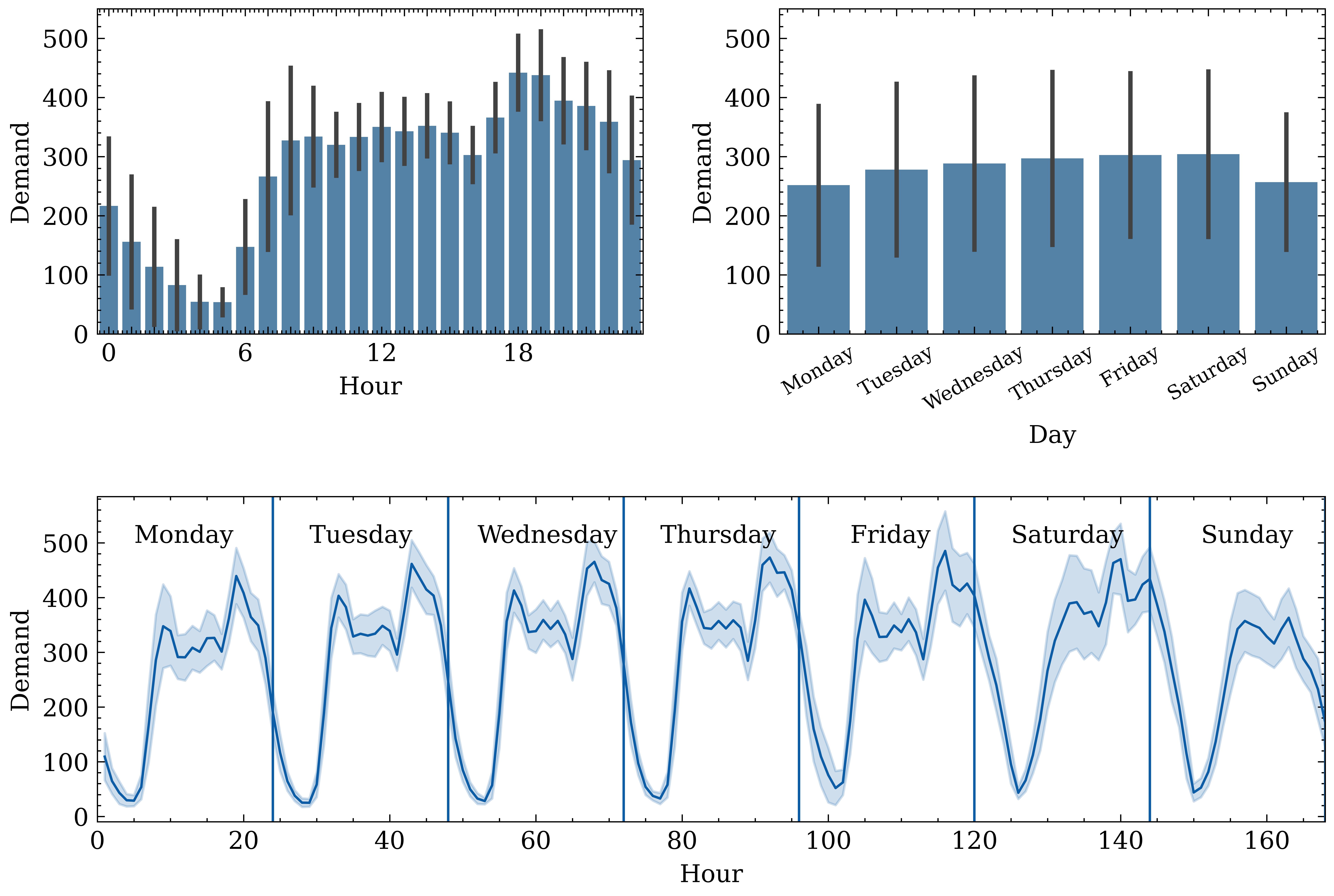}
	\caption{Temporal profiles of the demand dataset. We used 1.5 million trips recorded over 6 months of 2016 in Manhattan. The top figures present the average demand levels (trips per hour in Manhattan, which has an area of 60$km^2$) and their variability (standard deviation error bars) within-day (top left) and within-week (top right). The bottom row shows the average demand levels throughout the week and their standard deviation. Demand is distributed fairly uniformly on all days of the week, with a peak on weekend nights. Tuesdays to Thursdays are very similar, while Friday and Saturday are more pronounced at night, and Sunday has a substantially different profile. Demand typically reaches a flat plateau from 8 to 16 and peaks in the evenings. Weekends are much more variable in the demand.}
	\label{FIG:2}    
\end{figure*}

\begin{figure}
	\centering
		\includegraphics[width=\columnwidth]{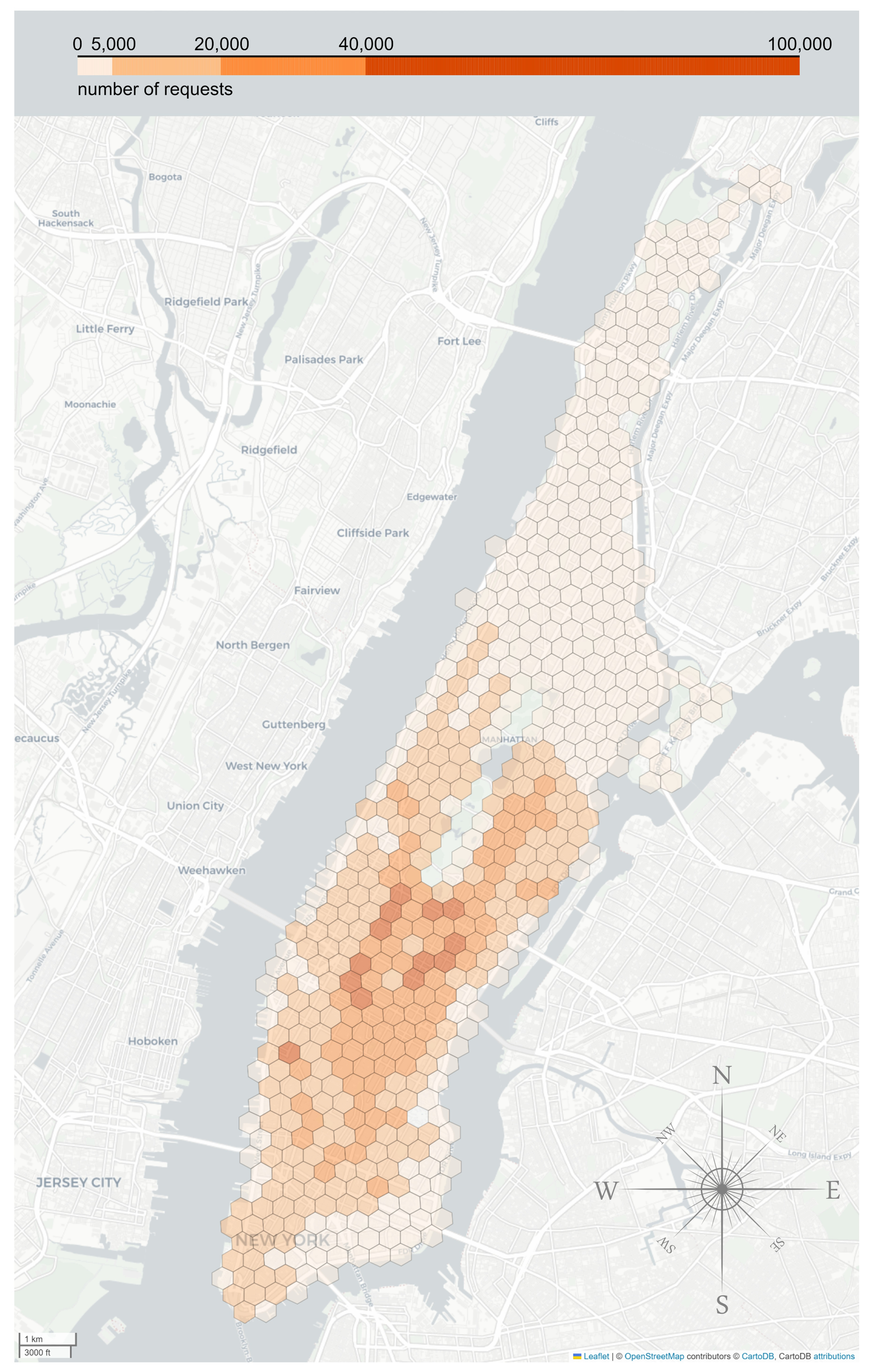}
	\caption{Spatial distribution of the trip origins, concentrated in the core and south parts and lower on the peripheries, with a decreasing trend from the central part of the network (Midtown) towards the north (Upper Manhattan).}
	\label{FIG:3}    
\end{figure}

\subsection{Solving ride-pooling problems with ExMAS}

We divide the dataset into about 9~000 half-hour  batches, for which we apply the ExMAS ride-pooling algorithm. ExMAS is an open-source Python algorithm that solves the offline ride-pooling problem for rational travellers \citep{kucharski2020exact}. For a given demand set of trips with their origin, destination, and departure time, ExMAS first identifies all the feasible (attractive) door-to-door pooled rides (groups of travellers that want to travel together) and optimally assigns each traveller to a vehicle with the objective of minimising total mileage. ExMAS is utility-based and assumes that pooling is selected only if it is attractive, i.e., rational travellers opt for pooled options for which the utility of pooling exceeds the utility of travelling solo. Utility is composed of travel time (weighted with value of time) and cost (per-kilometer fare multiplied with the distance). The fare is discounted if the ride is pooled, which shall at least compensate the detour (longer travel time),  delay (waiting for co-travellers) and discomfort (penalty multiplier for sharing). 
To assess whether a pooled ride candidate $r_k$ is attractive to the traveller $i$, we compare its perceived utility $U$ with the utility of a private ride by applying the following formulas:

\begin{equation}
\begin{aligned}\label{eq:utility_original_exmas}
& U^{ns}_i = \beta_c \lambda l_i + \beta_t t_i \\
& U^s_{i, r_k} = \beta_c (1 - \lambda_s)\lambda l_i + \beta_t \beta_s (\hat{t}_i + \beta_d \hat{t}^p_i) + \varepsilon,
\end{aligned}
\end{equation}

where $U^s_{i, r_k}$, $U^{ns}_i$ denote, respectively, the utility of shared ride (for $i$-the traveller, ride $r_k$) and the utility of non-shared ride (for $i$-th traveller). $\lambda$ stands for the per-  fare, while $\lambda_s$ denotes a discount for sharing a ride. $\beta^c$, $\beta^t$, $\beta^s$ $\beta^d$ are the exogenous behavioural parameters: cost sensitivity, value of time, willingness-to-share and delay sensitivity, respectively. $t_i$ and $\hat{t}_i$ stand for the travel time of non-shared and shared rides, respectively, $\hat{t}^p_i$ is a pick-up delay associated with pooling and $\varepsilon$ is a random term (for the sake of simplicity, assumed here to be constant and null, which yields a deterministic model). 
Thanks to such constraints, the algorithm identifies all the pooled rides of any degree for which the costs of pooling (detour, delay and discomfort) are at least compensated with reduced fare. ExMAS assumes that the ride is attractive if the utility of pooling is greater than the utility of travelling alone for all the travellers sharing a ride.

The final solution to the ride-pooling problem is found by solving the so-called matching problem. In the matching problem, ExMAS selects the subset of feasible rides that yield minimal total mileage, under the constraint that each traveller is uniquely assigned to a ride. Notably, the rides in the solution  may be pooled or solo (private ride-hailing, when no attractive matches are found). ExMAS is exact and explores all feasible rides of any degree (number of travellers), its results are consistent with discrete-choice theory; however, it assumes the demand is known in advance (offline), and the vehicle fleet is not explicitly modelled (demand-oriented). Thanks to such an approach, we can reveal the ride-pooling potential of historical trip records. 

The ExMAS is parameterized with traveller behaviour: value-of-time ($\beta_t$, which we assume to be 11,6~$\$/h$); penalty for sharing ( $\beta_s$, travel time multiplier due to discomfort of pooling, which we assume to be 1.3 based on the results of \cite{alonso2020value}. In the baseline scenario, we use the discount $\lambda$ of $32\%$ and then experiment with lower discounts of $20\%$, $24\%$, and $28\%$. We assume a trip fare $\beta_c$ of 1,38~$\$/km$ (according to \cite{uber2022}) and constant and network-wide flat speed of 21.6 $km/h$ (a very rough estimate for of highly variable actual network speeds in Manhattan, which, for sake of clarity, we did not incorporate in this longitudinal study). We use the standard waiting time multiplier of 2, per analogy with the public transport studies \cite{yap2020crowding}, and each pick-up drop-off operation takes 15s.

\subsection{Ride-pooling Performance Indicators}

We measure ride-pooling performance with the following set of indicators computed from the solution of ride-pooling problem for each of 9~000 batches in our case study:
\begin{itemize}
  \item Vehicle hours reduction ($\triangle{T}_{v}$) - measure benefits for service operator. It indicates the relative reduction of vehicle hours due to the pooling. Here, we calculate it as a relative difference between total vehicle hours (with travellers only) in the solo-ride scenario (when pooling is not available) and in our pooled scenario.
  \item Travellers utility gains ($\triangle{U}_{p}$) - measure benefits for travellers. It indicates the relative increase in travellers (dis)utility due to pooling, i.e. how well the discounted fare compensates the discomforts of pooling (detour and delay). Computed as the total utility gains for all the travellers being part of the ride-pooling problem solution, as expressed with eq.\ref{eq:utility_original_exmas}.
  \item Share of pooling (${S}$) - indicates a portion of travellers that managed to be successfully pooled into shared-rides, i.e. share of travellers which, in the final ride-pooling problem solution, have been assigned to the pooled rides.
  \item Occupancy (${O}$) - measures the pooling effectiveness. It indicates a ratio of the total passenger hours in the solo ride-hailing scenario to the total vehicle hours in the pooled scenario. It equals 1 in the solo scenario and reaches 1.6 in the most effective experiments.
  \item Passenger hours increase ($\triangle{T}_{p}$) - measures perceived costs for travellers. It indicates an increase in the total travel time of all travellers relative to the solo-rides scenario. 
  \item Potential fleet reduction ($\Delta F$) - roughly approximates the potential reductions in fleet size. Since vehicles are not treated explicitly in ExMAS, we approximate it using the maximum number of rides being undertaken at a given time. In the solo-rides scenario, this equals to the maximal number of simultaneous trips, while in the pooled scenario, this is reduced as some trips are pooled in the same vehicle. Measuring how pooling reduces the number of simultaneous rides offers a rough approximation, yet for exact fleet estimation this shall be refined with explicit fleet methods (like  \cite{alonso2017demand}).
\end{itemize}
To understand the impact of pooling to the above measures, we (following \cite{soza2022shareability}) compare vehicle hours, travellers utility, passenger hours, and fleet size when the ride-pooling service is available (subscripted $S$) and is not applied (subscripted $P$) as follows:

\begin{equation} \label{e2}
\begin{array}{ll}
\Delta T_{v} =\frac{\mathrm{T}_{v}^{S}-\mathrm{T}_{v}^{P}}{\mathrm{T}_{v}^{P}}; \quad \Delta U_{p}=\frac{\mathrm{U}_{p}^{S}-\mathrm{U}_{p}^{P}}{\mathrm{U}_{p}^{P}}; \\ \quad \Delta T_{p} =\frac{\mathrm{T}_{p}^{S}-\mathrm{T}_{p}^{P}}{\mathrm{T}_{p}^{P}};  \quad \Delta F =\frac{\mathrm{F}^{S}-\mathrm{F}^{P}}{\mathrm{F}^{P}} 
\end{array}
\end{equation}

\section{Results}
We start by confirming the notion of critical mass and its significance for ride-pooling performance in \fig{FIG:4}. On one hand we observe strong trends against all six  indicators, on the other the huge residual variability calls for more detailed analysis.
Then we illustrate distributions of six KPIs in \fig{FIG:5}~each following similar, yet subtly different profiles. In \fig{FIG:6}~we report the within-day variability of the results. Most of KPIs have similar within-day profile and variability, except travellers utility gains which remains flat during the day while others peak in the evenings. \fig{FIG:7}~illustrates the spatial concentration of pooled rides, with half of the rides concentrated in only 14\% of spatial hexagons. Spatial patterns of three selected indicators are presented in \fig{FIG:8}~with nuanced pictures. For a richer picture, we select one indicator and show its spatial distributions in eight different time periods in \fig{FIG:9}. In \fig{FIG:10}~we experiment to see how the price (discount) affects pooling performance. We conclude with smaller experiments in Chicago and DC (\fig{FIG:11}).

\subsection{Critical mass}
\CC{ \fig{FIG:4} ~reveals the strong, non-linear trend of demand  against all indicators. Each dot represents the result of respective KPI for a given demand in an experiment. The logarithmic curve (thick lines) fits the intuitive notion of critical mass well and fits the results of our experiments. 
We observe intensive growth at initial phase of low demand values and gradual stabilisation as the demand goes beyond the levels of 200 trips per hour. Stabilisation in all performance indicators indicates reaching the critical mass, at which the service becomes effective for all participants: for the platform, travellers and drivers. Notably, at high demand levels trends in all KPIs remain positive. }

The results remain highly variable, and the logarithmic trend against the demand levels fails to describe a residual variability. For the same demand level of 400 trip requests per hour, the vehicle hours reduction varies from 10\% to over 30\%; travellers gain between 2\% and 6\% of their perceived utility; 35\% to 70\% of trips are pooled; the occupancy varies from 1.2 to 1.6,  passenger travel times may increase either by 4\% or by over 10\%, and potential reduction in fleet size varies from 10\% to 40\%. 

\begin{figure*}
	\centering
		\includegraphics[width=\linewidth]{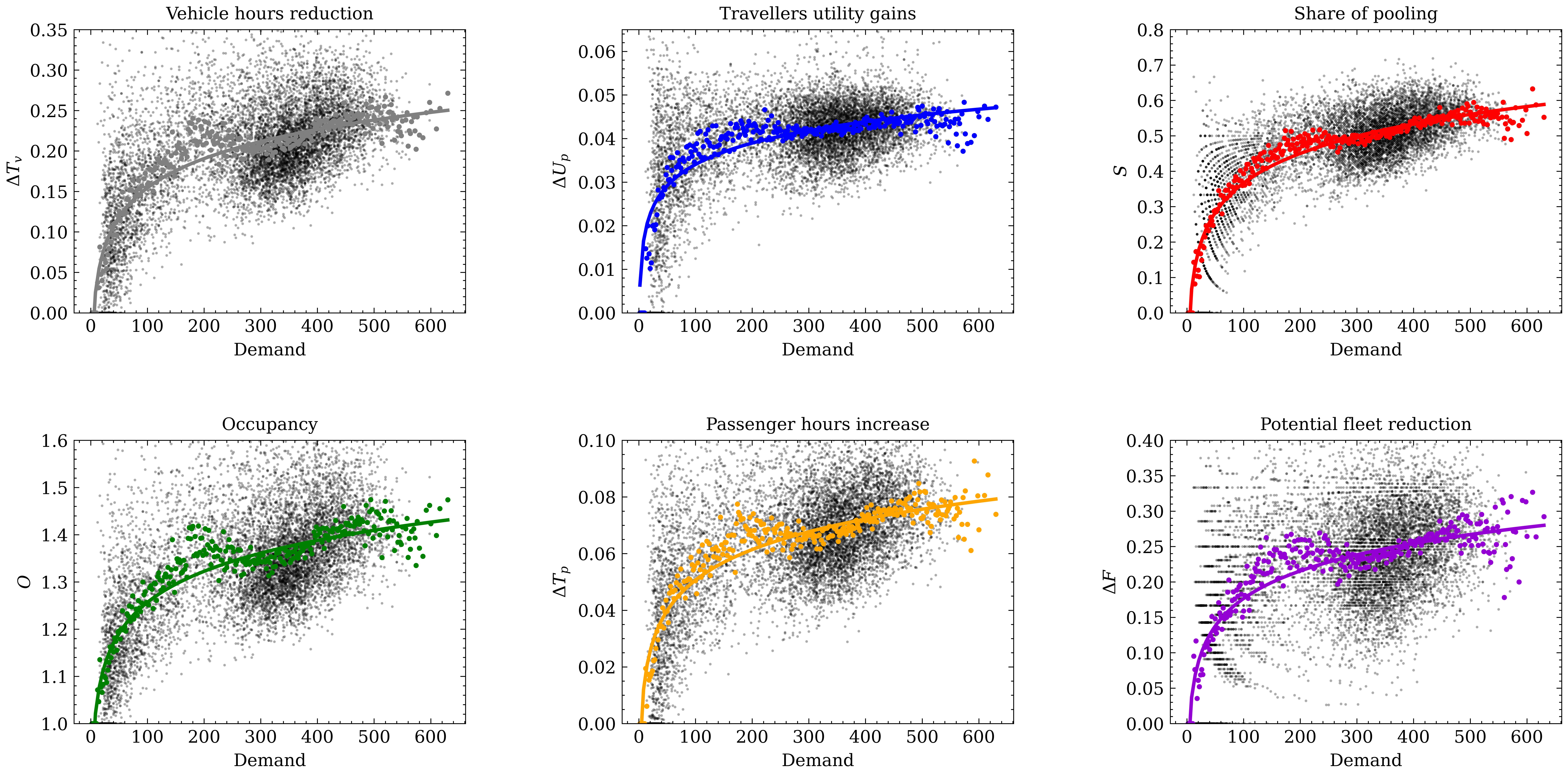}
	\caption{Six key performance indicators of ride-pooling plotted against the demand levels. Each dot represents the result of a single experiment (30-minute demand batch), thick dots denote average per demand level, and a thick lines denote a logarithmic trend fit. Each performance indicator follows a similar trend: it starts low, grows fast, and stabilises with a flat, yet still increasing trend for high demand levels.}
	\label{FIG:4}    
\end{figure*}

\subsection{Performance distributions}
 We follow with the distribution of each indicator in \fig{FIG:5}, each data point is a single batch experiment. Observations with null values show experiments in which none of the trips was successfully pooled (there were no batches with zero requests).
 While all the distributions are similar, there are differences.
 The distribution of utility gains has the smallest variance, while vehicle hours reductions and potential fleet reduction are most variable.
 Share of pooling has the longest left tail, while occupancy right tail. The share of pooling, passenger hours increase, and utility gains indicate the same (left-skewed) asymmetry with different magnitudes. Occupancy is symmetrically distributed, similarly to passenger hours increase, and potential fleet reduction, as well as travellers utility gains, which is also the most narrow of obtained distributions. The null values are obtained in ca. 200 batches, however, in 220 cases passenger hours did not increase at all (with over 400 batches with null fleet reductions on our proxy).
 

\begin{figure*}
	\centering
		\includegraphics[width=\linewidth]{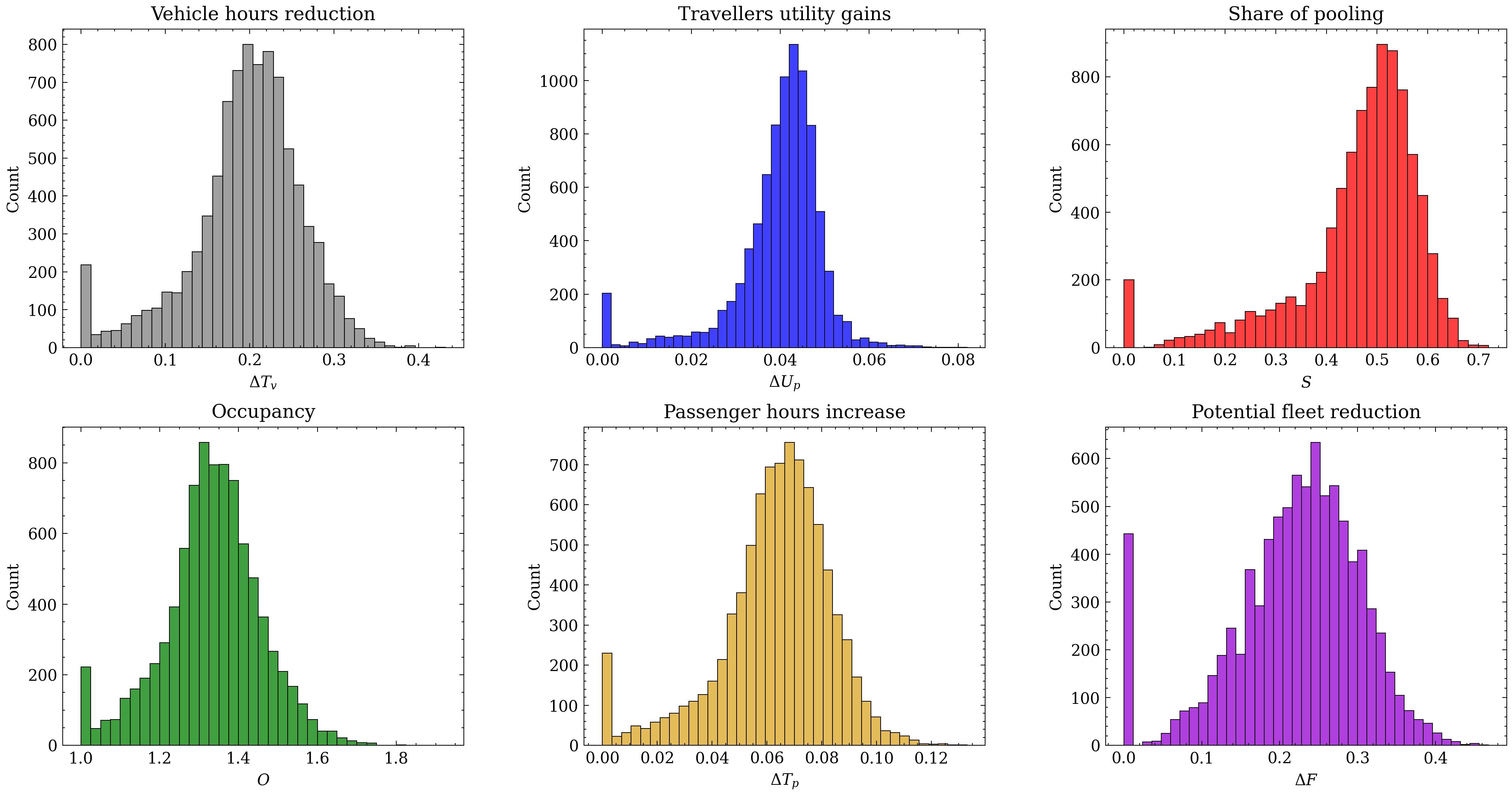}
	\caption{Distribution of observed key indicators of ride-pooling. Each data-point is the result of one ride-pooling experiments on a 30-minute batch. A significant share of null observations (where sharing was not induced at all) was reported for all six indicators. The remainder of the distribution follows various shapes: symmetrical (like occupancy), left-skewed with fat left tail (like share of pooling and passenger hour increase) or narrow (like utility gains).}
	\label{FIG:5}    
\end{figure*}

\subsection{Within-day performance}

Within day mean and standard deviation of indicators is shown in \fig{FIG:6}. 
A fairly large variability in the values of the analysed indicators suggests the presence of other factors, apart from temporal, that affect the ride pooling performance. While the profiles of average values are similar throughout the day, the variability of utility gains is the greatest, in contrast to the smallest variability of its distribution. Notably, distributions are different from the demand pattern (\fig{FIG:2}), utility gains do not increase in the evening and remains reasonably flat throughout the day, while the increase in the evenings' occupancy and vehicle hours reductions is significant. 


\begin{figure*}
	\centering
		\includegraphics[width=\linewidth]{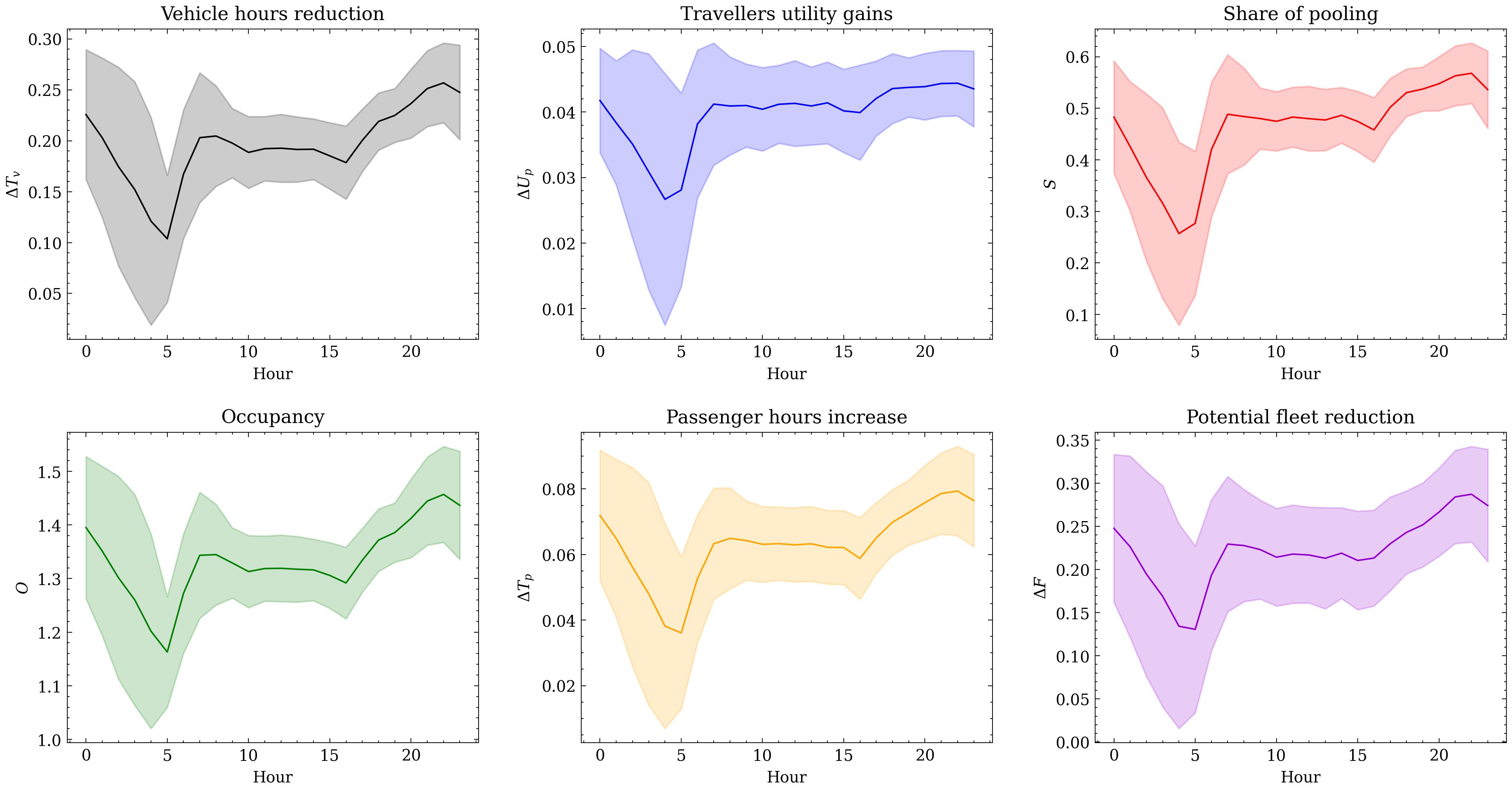}
	\caption{Within-day ride-pooling performance. Averages of six indicators observed throughout the day (thick lines) and their standard deviations. Four indicators have remarkably similar profiles with low performance and high variability at night, a flat plateau during the day, and a peak of performance in the evening. Only travellers utility gains follow a slightly different pattern, with less significant increase in the evening (despite high occupancy and vehicle hours reductions).}
	\label{FIG:6}    
\end{figure*}

\subsection{Pooling concentration}

In \fig{FIG:7}, to understand how the pooling is distributed across Manhattan, we show the concentration of pooled trips in the space (hexagons). 
Pooled trips origins are highly concentrated, with 50\% of them covering only 14\% of spatial hexagons. 
For this and further spatial analyses, we divided the area into 532 hexagons (using the Uber's H3 library \cite{Brodsky2018}), each of 532 hexagons has an edge length of 200m and area of over $105~000m^2$.

\begin{figure}
	\centering
		\includegraphics[width=\columnwidth]{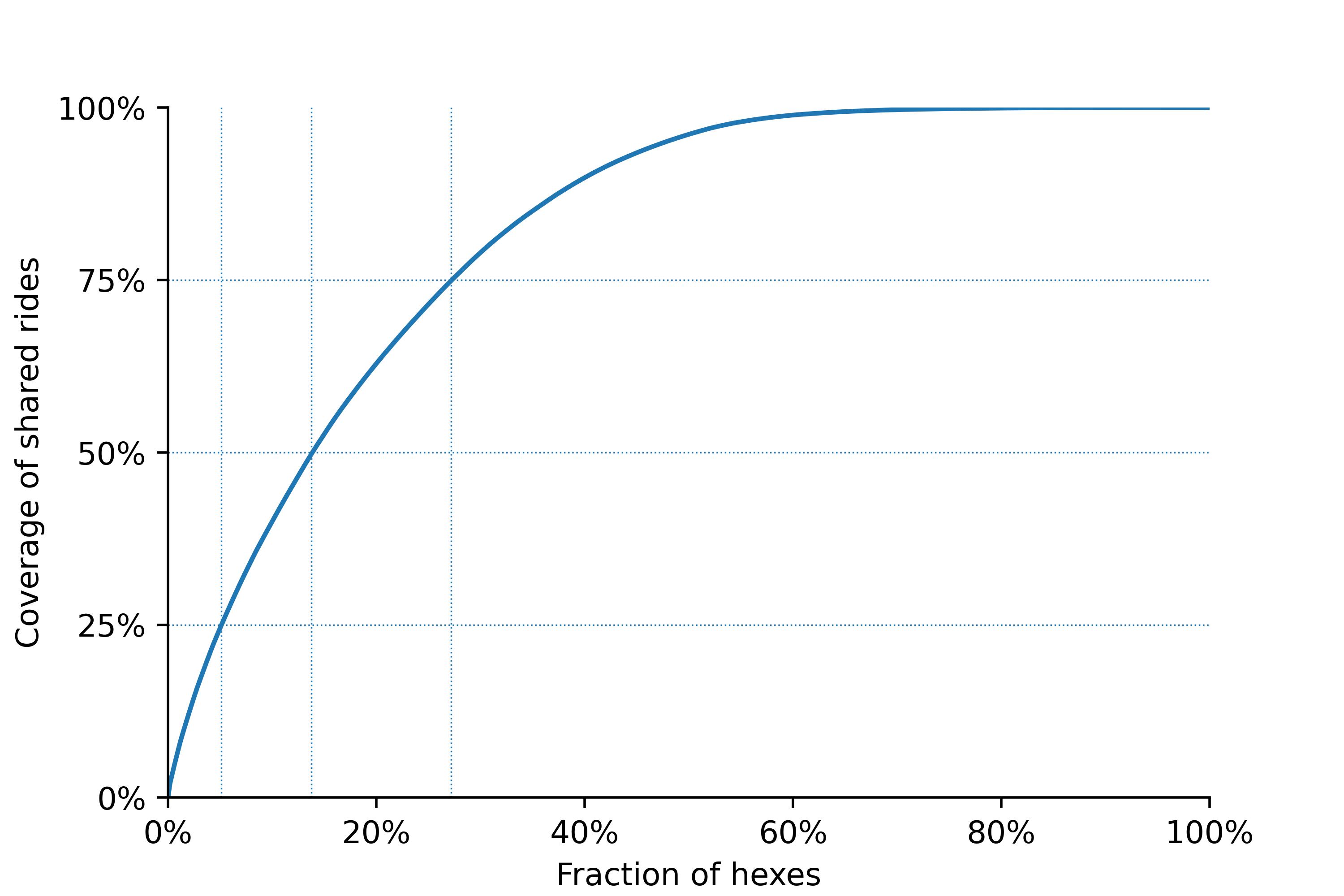}
	\caption{Concentration of pooled trips origins in the space. The top $13.8\%$ of hexes account for half of all pooled rides, with $5\%$ of hexes containing $25\%$ of the rides and $27.2\%$ covering $75\%$. It indicates an unevenly distributed demand for ride-pooling and the presence of territorial centres where ride-pooling is the most attractive.}
	\label{FIG:7}    
\end{figure}

\subsection{Spatial patterns of pooling performance}
We plot the spatial distributions for three indicators: travellers utility gains, share of pooling and increase in passenger hours in \fig{FIG:8}. We used the requests' origins as a spatial reference and aggregated mean values over spatial hexagons. Since occupancy and vehicle hours reduction is calculated not for requests (single trips) but for rides (of presumably multiple passengers with multiple origins), mapping them spatially is ambiguous and we refrained from plotting it.

Spatial distributions allowed us to reveal interesting patterns. We expected the maps to simply resemble the demand distribution (\fig{FIG:2}) as the demand density is the main driver of pooling performance. Surprisingly, the patterns are more nuanced. Utility gains (\fig{FIG:8}a) are concentrated in the central part of Manhattan, spanning to its southern tip. The share of pooled rides has a smaller concentration area, confined to the central part, yet with another hotspot in the southern part of Manhattan (\fig{FIG:8}b). The pattern of detours is substantially different, peaking in the south, with few smaller peripheral areas with higher detours (\fig{FIG:8}c). 
Despite the clear patterns and big dataset, several outliers appear, typically in the upper Manhattan (North).

\begin{figure*}
	\centering
		\includegraphics[width=\linewidth]{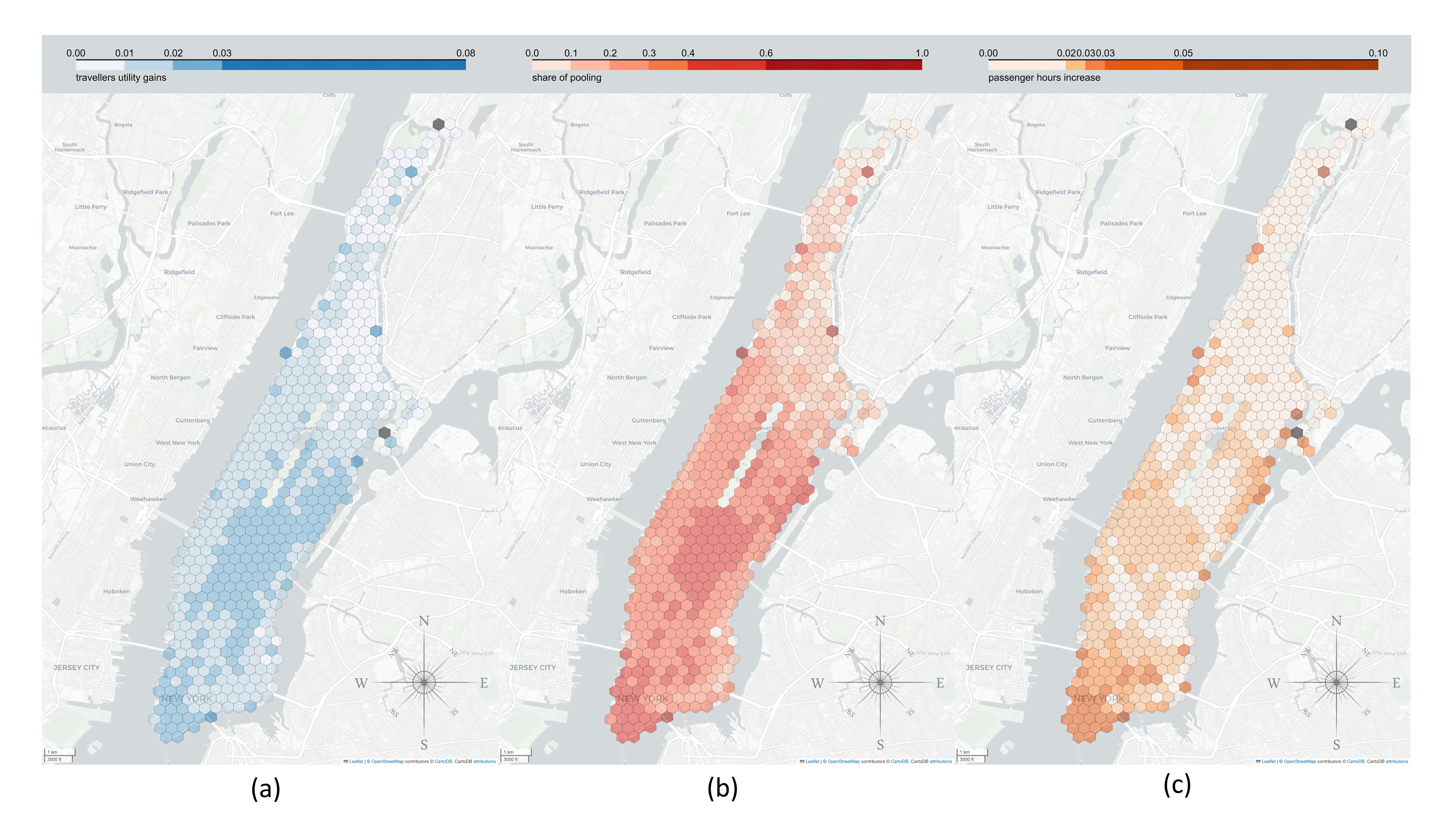}
	\caption{Spatial patterns of potential ride-pooling performance at trip origins. Travellers utility gains (a) are pronounced in the central part of the network, but the greatest share of pooling (b) is observed also in the southern part (Wall Street), here, however, the increase in passenger hours is the greatest (c). }
	\label{FIG:8}    
\end{figure*}

We complement the above with spatiotemporal analysis. We report in \fig{FIG:9} ~how the spatial patterns of share of pooled rides varies within the weekday (top) and weekends (bottom) in the: morning, afternoon, evening and night. 

Weekday mornings and afternoons are somehow complementary, as the morning origins often become afternoon destinations. The evening pattern, with bigger shares of pooled rides in general, is more uniformly distributed, both in weekdays and weekends. 
While weekend afternoon has the similar pattern to weekday, weekend mornings are different, more uniform. 
A more uniform pattern may explain high pooling performance in the evenings. 

\begin{figure*}
	\centering
		\includegraphics[width=\linewidth]{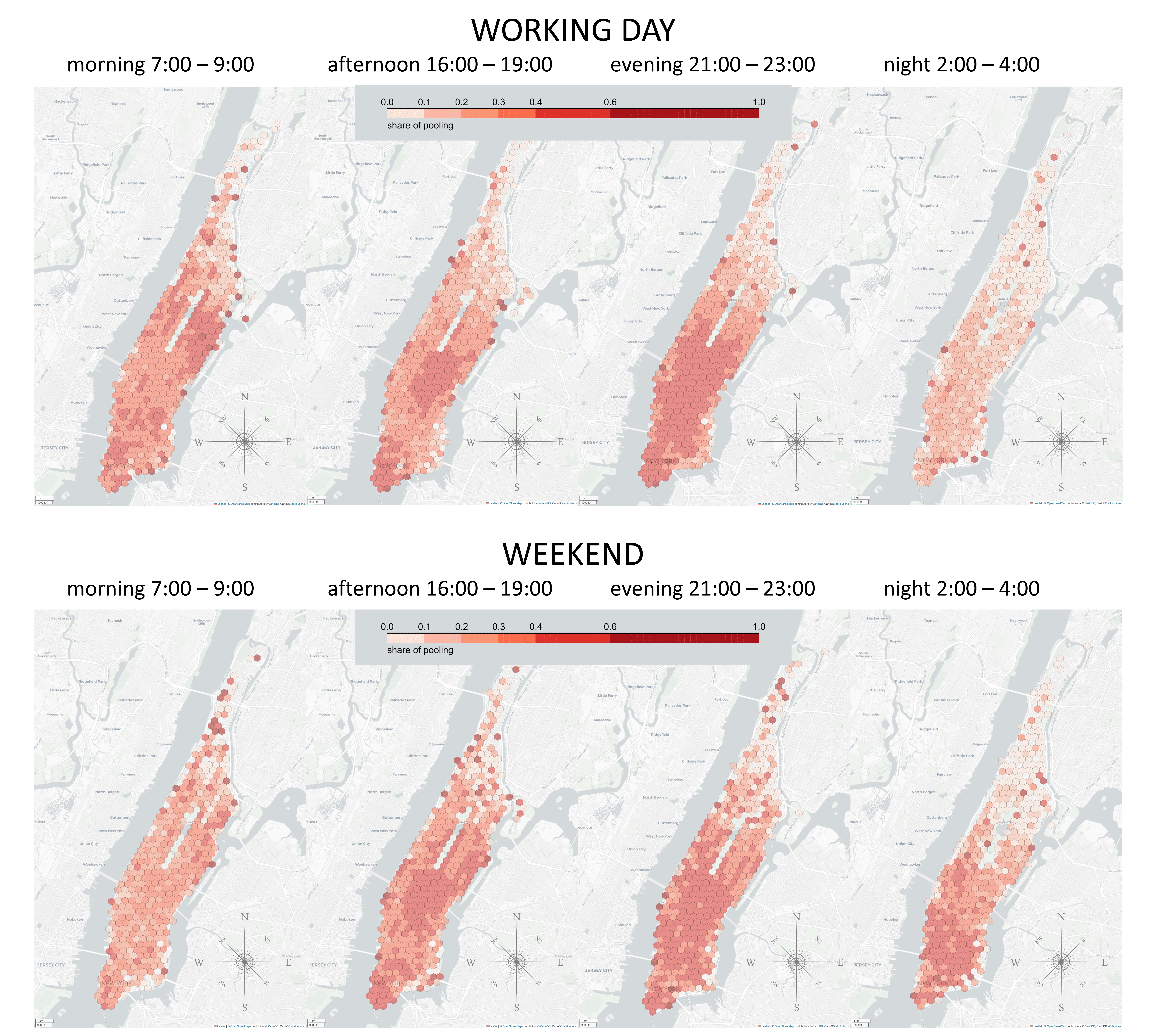}
	\caption{Spatial patterns of the resulting share of pooled rides for the ride-pooling problems of the working days (top) and weekends (bottom) for four periods of the day (columns). In the afternoons, evenings, and nights of the weekend, half of the trips can be attractively shared, while in the evening the highest concentration is uniformly spanned over Manhattan. During the day there is a concentration in the central and the southern parts, and at night in the part of Manhattan between the central and southern parts.} 
	\label{FIG:9}    
\end{figure*}

\subsection{Controlling performance with a discount}

We leverage on our utility-based approach and test travellers' sensitivity to the pooling incentives. We see how the four various discounts offered for ride-pooling (relative to the ride-hailing fare) attract travellers. 
While previous results were computed for the 32\% discount, now we experiment with values of 20\%, 24\% and 28\% discounts. 
This affects the system performance, as we demonstrate in \fig{FIG:10}. Notably, the demand levels in those experiments remain intact, and the discount only affects how many travellers are satisfied and opt for pooling (as expressed with eq. \ref{eq:utility_original_exmas}).

With 32\% discount for 200 out of 9~000 batches there was null potential to pool. This increases to over 700 batches when the service provider offers 20\% discount (\fig{FIG:10}~right). The average occupancy decreases when decreasing the discount; it shifts from 1.35 to 1.1 with the 20\% discount. Notably, for the service to be profitable, the discounts (which are perceived by the platform as costs) need to be compensated with benefits (which can be indirectly proxied with occupancy). This means that offering too low discounts will not induce enough benefits, while offering too high discounts will reduce the costs beyond the obtained benefits. We argue that offering 20\% discount in the above experimental setting is insufficient to induce sustainable pooling, while 32\% discount attracts the travellers to pool, yet the vehicle hours are reduced by only 13.7\%.

\begin{figure*}
	\centering
		\includegraphics[scale=0.8]{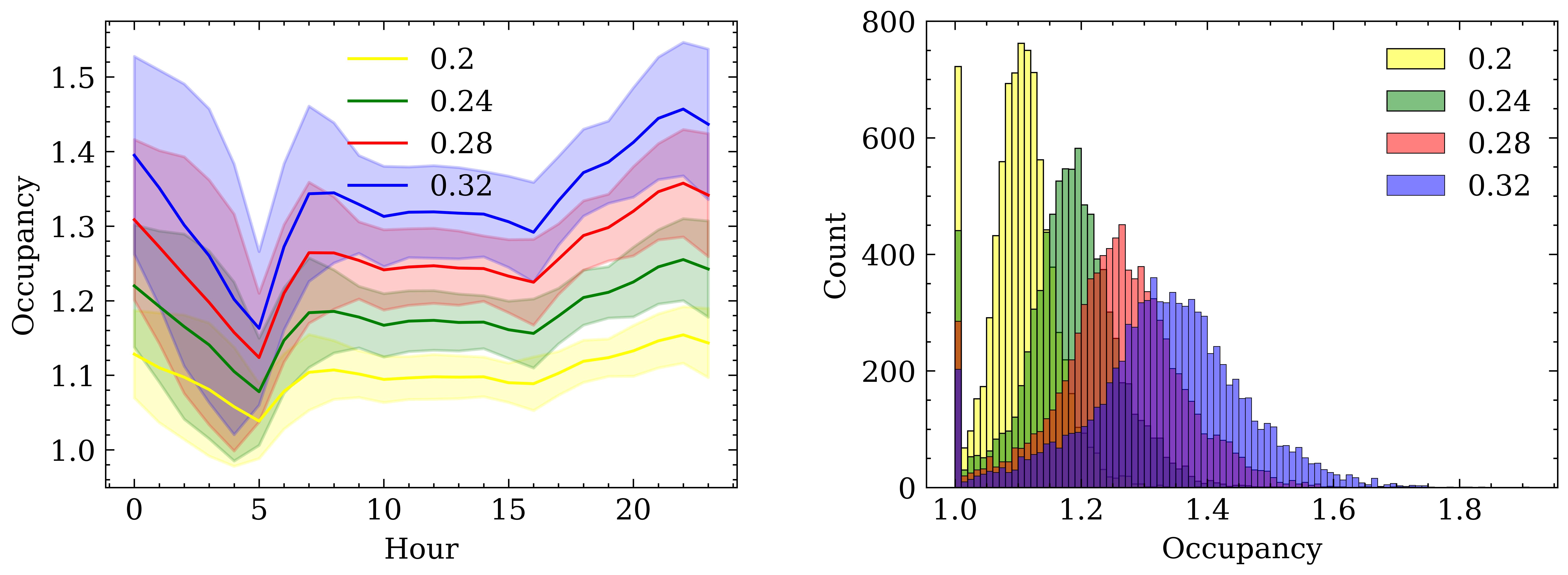}
	\caption{Controlling the ride-pooling performance with the ride-pooling discount (relative to ride-hailing). The average occupancy (central ride-pooling efficiency indicator) increases with the discounts offered, both within the day (left) and in general (right). The within-day profiles have the same shape, yet magnified. The share of solo rides (bars at occupancy of 1 in the right panel) substantially decreases with increasing discount, and the occupancy shifts to the right.}
	\label{FIG:10}    
\end{figure*}

\subsection{Validation in Chicago and Washington, DC}

To see if the detailed NYC findings would generalise to other cities, we run smaller experiments with the two cities for which similar datasets were available: Chicago (from February 2023, 641 half-hour batches) and Washington (from September 2019, 1468 batches). Despite the unique spatial setting of Manhattan, with flattened and dense demand patterns, the occupancy (most important KPI in our analysis) followed very similar growth patterns (\fig{FIG:11}), both in terms of shape, values and variability. This, coupled with other studies suggesting generalisation of ride-pooling performance \citep{tachet2017scaling} can be used as an argument for some universality across different topologies and land-use patterns.

\begin{figure}
	\centering
		\includegraphics[width=\columnwidth]{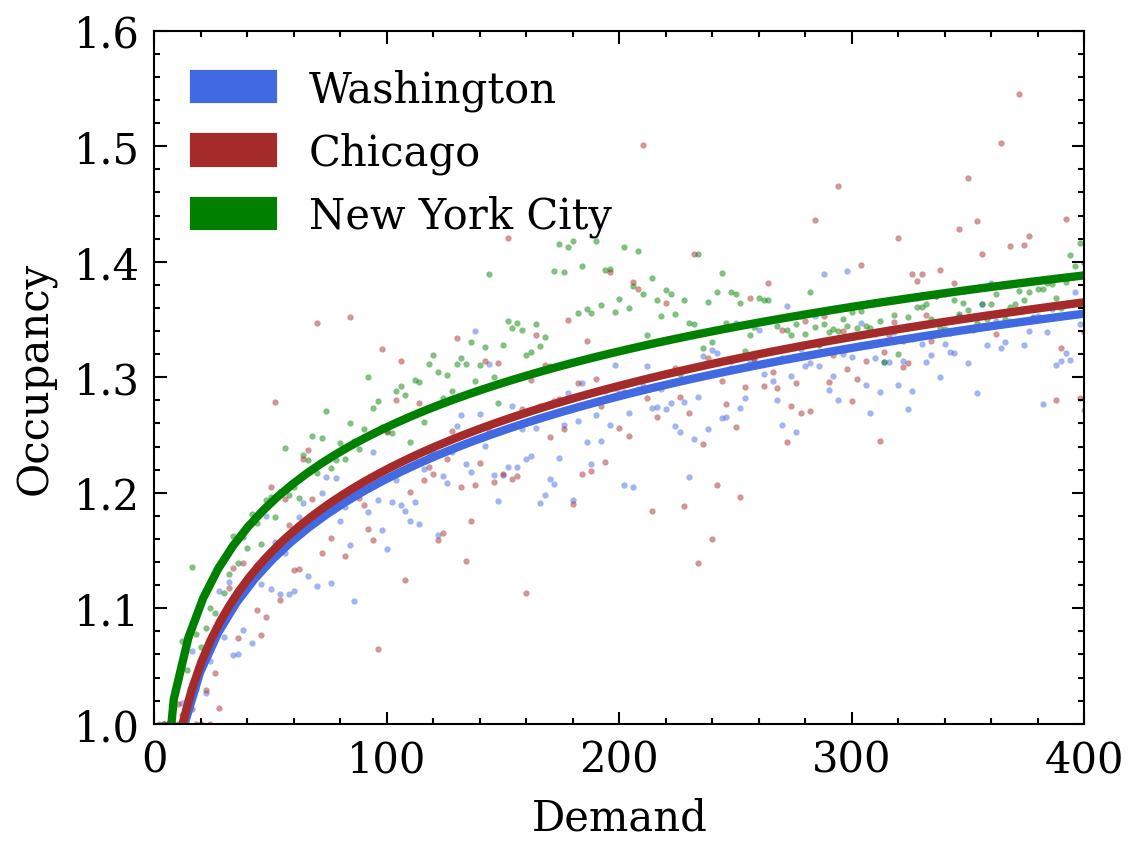}
	\caption{\color{black}Generalisation of the occupancy results to two other cities: Chicago (ca. 600 experiments) and Washington (ca. 1500 experiments). Despite highly different urban structure, land-use and topology, occupancy for all three cities follows both similar values (reaching average occupancies of 1.4) and trends (starts low, grows fast, and stabilises with a flat, yet still increasing trend for high demand levels).\color{black}}
	\label{FIG:11}    
\end{figure}

\section{Conclusions}

Revealing spatiotemporal patterns of the system performance allowed us to better understand the potential of ride-pooling services. By experimenting with the utility driven ride-pooling algorithm on the big set of actual ride-hailing trips in New York, we obtain the rich picture of ride-pooling potential by reporting six different performance indicators: vehicle-hour reductions, travellers' utility gains, share of pooling, occupancy, passenger hours increase, and potential fleet reduction. They reveal the complex and nuanced spatial and temporal patterns of potential pooling performance in the first half of 2016 in Manhattan.

If offered with a 32\% discount, around 695~000 out of 1.5 million trip requests would be successfully pooled into attractive rides. This would reduce vehicle hours by 25.2\% and improve the passengers' perceived utility by 4.1\%. The vehicle occupancy could reach up to 1.8 passenger hours per vehicle hour, with the mean at 1.33. The fleet size could be reduced by 40\% in the most effective periods, with the mean reductions at 25\% (\fig{FIG:4}). Only in 2.3\% of the cases the pooling would fail to attract at least one traveller, and in 45.3\% of the analysed half-hour scenarios more than half would pool} (\fig{FIG:5}).
The potential is greatest in the evenings; at 10PM the vehicle hours could be saved by 25\% on average and only by 10\% at 5AM (\fig{FIG:6}). The evenings, presumably due to their unique spatial patterns, break out from the correlation with the demand: the demand decreases from 6 to 11PM and the ride-pooling performance increases. The expected occupancy reaches 1.45 at 10PM and drops to 1.35 during the day (8AM-4PM).

We argue that while demand level well represents the general patterns, temporal and spatial components are crucial to better understand the ride-pooling potential (\fig{FIG:4}). With a demand of up to 150 travellers per hour, intensive growth is observed in all six indicators. After reaching the critical mass, increasing demand improves the ride-pooling performance (vehicle hours reduction) from the platform’s perspective, but its impact on travellers utility is lower.  Notably, unlike the classic works of \cite{tachet2017scaling}, all KPIs increase also at high demand levels.
For all indicators in the demand stabilisation stage the variability decreases and the single observations are concentrated around mean. 
Interestingly, the expected travellers utility gains and share of pooling remain almost stable in the range of 250 to 450 travellers per hour (per 60$km^2$).

We reveal the distributions of indicators obtained over the half-year study 
(\fig{FIG:5}) with similar, yet subtly different shapes. 
The average occupancy in the six-month experiment was around 1.4, yet we observed values ranging from occupancy of 1 (no pooling) up to 1.8. Suggesting the existence of time periods where pooling is highly effective, as well as periods without any potential to pool. 
The temporal profiles of six KPIs follow very similar, yet subtly different trends, both in means and in variances (\fig{FIG:6}), strongly correlating with the demand. The mean occupancy in the evenings is above 1.45 and drops to 1.2 during the night (\fig{FIG:6}). Low demand from midnight to early morning yields less efficient ride-sharing, two peak hours during the day (7:00AM and 9:00PM) lead to high system performance. Notably, while the average demand decreases from 6PM to 11PM (\fig{FIG:2}) the performance over that period improves in all KPIs.

The shared rides are highly concentrated in space (\fig{FIG:7}), $75\%$ of all pooled rides origins are in 27\% of spatial hexagons into which we divided Manhattan. 
Spatial patterns do not necessarily follow the demand density. More nuanced patterns are revealed. Travellers' gains are concentrated in the central part of Manhattan (\fig{FIG:8}a), the share of pooling is high also at the southernmost tip (\fig{FIG:8}b). Surprisingly, the detours are greatest at the southern tip and are scattered around other parts of the city (\fig{FIG:8}c). The results in \fig{FIG:8} can be, to some extent, explained with the demand distribution in \fig{FIG:3}. 
In high demand areas travellers utility gains are high because of many travellers to potentially share one's ride. But, interestingly utility gain for travellers is highest at southernmost tip, presumably because they are longer (in longer trips passengers benefit from ride sharing more and better compensate the detours and delays).

We show spatial patterns of share of rides that were actually pooled for workdays and weekends in the morning, afternoon, evening and nighttime intervals (\fig{FIG:9}). In the afternoons, evenings, and nights of the weekend half of the trips can be attractively shared, while in the evening the highest concentration occurs in almost the entire territory of Manhattan except for the northern part. For the weekend mornings, in many cases less than $40\%$ of trips are shared, in the weekday nights often times no trips are pooled at all (\fig{FIG:9}). The difference in those patterns may explain high pooling performance in the evenings, despite lower demand levels, the demand distribution is more uniform in the evenings and rides are pooled equally across the core of Manhattan. 

Thanks to our utility-based approach, the travellers' satisfaction is (to some extent) controllable by the discount offered for pooling. By increasing discount from $20\%$ to $32\%$ the number of cases when no pooling is observed, drops over threefold and the average occupancy raises from 1.1 to 1.35 (\fig{FIG:10}). 
Our validation in Chicago and DC is promising, since we observe very similar trends in the occupancy. In future, a more detailed spatiotemporal analysis of those datasets shall determine if our findings are general.

\section{Discussion}

Ride-pooling still has the potential to become an attractive alternative mode of transport, presumably reducing car dependence and vehicle mileage in urban areas. 
To realize this potential, we need to understand it in depth. 
Demand density is not the only factor driving the ride-pooling potential. As we demonstrate, for the same demand levels, the efficiency of pooling services may vary significantly. 
Here, we tried to understand this residual variance with the longitudinal experiment with big data in New York City.

The demand levels greater than 180 trip requests per hour in the area of Manhattan ($60km^2$), i.e., 3 trip requests per hour per square kilometre, allow reaching critical mass of ride-pooling in Manhattan, with ca. 20\% reductions in vehicle hours, travellers' utility improved by 3.5\% compared to solo ride-hailing and over 40\% of trips actually pooled into attractive rides of average occupancy greater than 1.3 passenger hour per vehicle hour. When demand doubles (to ca. 6$trips/h/km^2$) performance improves, but not drastically (50\% of pooled rides and occupancy of 1.4). However, this is far not guaranteed, as 6$trips/h/km^2$ may yield 13 or 33\% vehicle hour reductions (\fig{FIG:4}).

Pooling is most promising in the core and less in peripheries, at peripheries, even when pooling is attractive its costs are higher as it yields longer detours there. The greatest potential area shifts from the North in the morning to the Central and South in the afternoon. Weekend mornings have low and uniformly distributed pooling potential, which concentrates in the South of Manhattan on weekend nights.
Pooling in the evening is most efficient, presumably is due to uniformly distributed share of pooling (\fig{FIG:9}), unlike higher, yet more concentrated demand in the afternoons.

The 32\% flat discount seems to be sufficient to attract pooling from ride-hailing. We argue that it is a good balance: lower discounts make pooling less attractive and thus less used, while greater discounts would induce longer detours, which could not be compensated for the platform. Whether such services need to, or should be, publicly subsidized remains opened. Here, we show that the 32\% fare reduction is compensated with only 25\% reductions in vehicle hours, which may deem unprofitable for the commercial operator. This is inline with observation of policies of the greatest ride-hailing platforms, who seem to lose enthusiasm to offer pooling in the post-pandemic era.

Ride-pooling is not colinear in its performance indicators. As depicted in \fig{FIG:8}, the maximum mileage reduction is not necessarily associated with the maximised travellers utility or the share ratio. It implies that policymakers and researches should be clear about their objectives. In particular, advocating ride-pooling benefits by maximising the sharing ratio only seems to be an inadequate approach.

The practical implications of this analysis shed light on the future strategy of introducing pooled services. Specifically, the proposed post-pandemic Uber pooled services \citep{Tsay2023} with a 5\% to 20\% discount seems to be not attractive enough to become successful. However, pooling services may benefit when tailored to the demand conditions. 
Our results would suggest offering pooled rides: in weekday afternoons from South and Central Manhattan, in the evenings without spatial limitation, and in the mornings from Northern Manhattan. 
Pooling shall not be offered at night, as it is neither efficient for the platform, nor attractive for travellers. The discounts may be time and space varying, e.g. 28\% discount in the evening and 32\% during the day yields the same occupancy levels. The discount may be lower in the core of the network, where disutilities of pooling (detours) are lowest and benefits highest.

Policymakers can exploit the revealed spatiotemporal patterns of pooling performance, e.g. by forcing ride-hailing companies to offer ride-pooling in the periods and areas when it is attractive for the city (vehicle hour reductions). Or subsidise the pooling in the less attractive periods and areas to improve the accessibility. This shall be controlled with the total mileage, which shall not increase after introducing ride-pooling services. In the event of the new pandemic, we may now understand when and where the ride-pooling services may be proposed as the intermediate transport mode between private cars and mass transit. With occupancy being a proxy for virus spreading, one can decide which spatiotemporal patterns of NYC trips may be served with ride-pooling, to balance between system's performance and virus-spreading.



Despite its merits, this study has certain limitations. 
First, it uses the revealed demand of taxi users, which may have substantially different spatial and temporal structure than the general demand.
Second, the behavioural assumptions used in this study can be extended to cover heterogeneity of ride-pooling aptitudes, as revealed, e.g., in \cite{alonso2020value}, where the population varies from enthusiasts to pooling-averse. 
The ExMAS algorithm is limited to door-to-door ride-hailing, the attractiveness is evaluated against solo ride-hailing only, the demand is known in advance and fleet is not explicit.
We argue that our approach with implicit fleet operations is not a limitation. 
In general, the fleet needed to operate the pooled rides shall not be greater than for solo ride-hailing. Thus, on average, the waiting times for pooled rides, with the same fleet of vehicles, shall be lower than for ride-haling. 
Finally, in this study we used network-wide and constant speed, which is obviously wrong for the congested cities and may affect both the results from the congested periods (AM peak) where due to slower trips, detours are longer and pooling would become less attractive, and, inversely, for uncongested periods (nights), where faster trips may make the longer detours more acceptable.

In future, this methodology may be applied to more cities, like Chicago and DC for which we ran small validations, with elastic demand and real network speeds. This would allow drawing more generally applicable conclusions and presumably reveal universal patterns of shareability. 



\section{ Credit authorship contribution statement}
\textbf{Olha Shulika}: Investigation, Writing - Original Draft. 
\textbf{Michal Bujak}: Investigation.
\textbf{Farnoud Ghasemi}: Visualization.
\textbf{Rafal Kucharski}: Supervision, Conceptualization, Writing - Reviewing and Editing. 

\section{Declaration of Competing Interest}
On behalf of all authors, the corresponding author states that there is no conflict of interest.

\section{Acknowledgements}
This research was funded by the National Science Centre in Poland program OPUS 19 (Grant Number \\  2020/37/B/HS4/01847).

\printcredits

\bibliographystyle{cas-model2-names}

\bibliography{ref1}


\end{document}